# Robust Chaos Generation on the Basis of Symmetry Violations in Attractors

Evgeny Nikulchev
Moscow Technological Institute
Moscow, Russia, 119334, Leninckiy pr., 39A
nikulchev@mail.ru

*Abstract*— We present a method for generating robust chaos. It is based on the search algorithm weak symmetry violation in the reconstructed attractor. On its basis the smooth functions in the form of a system of finite-difference equations. To ensure robust chaos generator introduced piecewise continuous member. The simulation results are given in the report.

*Index Terms*— chaos, robust chaos, symmetry.

## I. Introduction

Now chaotic models are used in very different areas - in metallurgy, machine-building, automation of technological processes in the financial sphere, in geology, medicine, for modeling of traffic in computer networks, security in corporate networks and cloud infrastructures, agriculture and many others [1]. This shows the great interdisciplinary value of scientific results in the theory of chaotic dynamics.

Currently, there are three scenarios of transition to chaos:
- Feigenbaum scenario (through a cascade of period-doubling bifurcations).
- Afraimovich–Shilnikov scenario (through the destruction torus or through the destruction of a closed invariant curve).
- Pomo-Manneville scenario (alternating in time almost regular oscillations with intervals of chaotic behavior, which is observed immediately after the threshold of chaos).

Study on scenarios of transition to chaos plays an important role in practice, since in some cases allow you to explore the onset of chaotic regime behavior of a dynamical system by changing parameters. Scenarios describe the behavior of the system near the critical value of the control parameter when the evolution of the system is not yet completely messy, but not quite regular.

In this article we consider a system in which there is a steady expectation chaos [2].

It is assumed that the time series is observed result the operation of some a discrete

$$\mathbf{x}_{k+1} = f(\mathbf{x}_k, \mathbf{x}_0), \qquad (1)$$

or a continuous system

$$\frac{d\mathbf{x}(t)}{dt} = F(\mathbf{x}(t), \mathbf{x}(0)) . \qquad (2)$$

Here $\mathbf{x} = (x_1(t), ...., x_n(t))$; $n$ — dimension of the phase space; $t$ — time; $k$ — discrete time (number); ($F, f$ — the unknown vector function).

The task is to determine the functions of the output (measured) value.

For systems with chaotic dynamics, there are methods to build an attractor from the observed series. The methods for evaluating the characteristics of numerical chaos time series.

The developed simulation of chaotic systems based on analysis of the attractor. The paper describes a method and shown that it can be used to generate robust chaos.

## II. Search Symmetry Violation

This is a essence of the developed modeling approach, based on the finding of a weak symmetry breaking [3, 4]. In input — time series, in output — model in the form of finite difference equations

- Calculated by numerical methods necessary conditions for the existence of chaos — the largest Lyapunov exponent (chaotic dynamics must be greater than zero).
- Reconstruction attractor.
- The assumption of chaotic dynamics.
- Searched the symmetry group (in a weak symmetry violation)
- According to the adopted symmetries Hausdorff formula, we obtain the form of equations in a minimal invariant manifold
- Structure equations parametrically identified
- Ranked dynamics model and the original series
- Calculated value of the forecast horizon (the adequate operation of the generator)

Development algorithm of the search symmetry violation [5]:

**1. Marking**. Isolation monotone fragments with only local extremes in the area.

**2. Normalization.** Converting part of a descriptor - the image that is invariant under the transfer, rotation and scaling

of the original title, as well as obtaining the numerical parameters of this transformation.

**3. Evaluation symmetry breaking.** After receiving two descriptors presumably symmetrical fragments and can give a numerical estimate of divergence (distance) between them, it is proposed to use the properties of the Fourier series.

**4. Genetic algorithm to search the entire trajectory.**
Criteria for finding solutions has the form:

$$D_{A,B} = \sum_{i=1}^{q}\left((I_{AB} + R_{AB})\cdot \beta_i\right),$$

$$I_{AB} = \sqrt{\sum_{j=1}^{n}\left(\text{Im}(S_{B,i,j}) - \text{Im}(S_{A,i,j})\right)^2},$$

$$R_{AB} = \sqrt{\sum_{j=1}^{n}\left(\text{Re}(S_{B,i,j}) - \text{Re}(S_{A,i,j})\right)^2},$$

where $q$ — number of conjugate pairs of elements of the spectrum, $n$ — dimension of the phase space, $\beta_i$ — coefficients ($i = \overline{1,q}$), which determine the degree of importance of the spectral components at frequencies, $S_{(\cdot)}$ — corresponding elements of the spectrum circuits $A$ and $B$.

### III. COMPUTING EXPERIMENT

Example 1. Use for computing the test time series shown in Fig. 1.

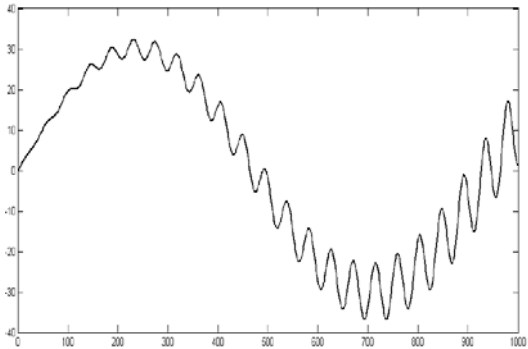

Fig. 1. Time series from example 1

It was reconstructed attractor. Attractor markup set in accordance with the developed algorithm is shown in Fig. 2. Genetic algorithm found almost symmetrical parts shown in Fig. 3.

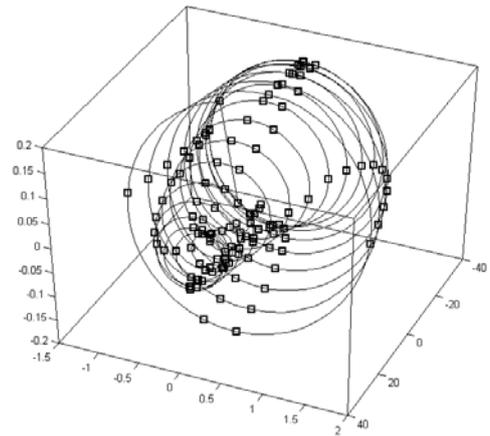

Fig. 2. Tagged Reconstructed attractor

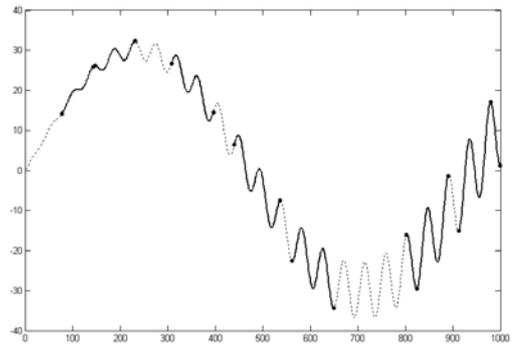

Fig. 3. Almost symmetrical parts

Example 2. Traffic network. The simulation was performed for solving traffic management. Symmetrical sections in small disturbances are shown in Fig. 4.

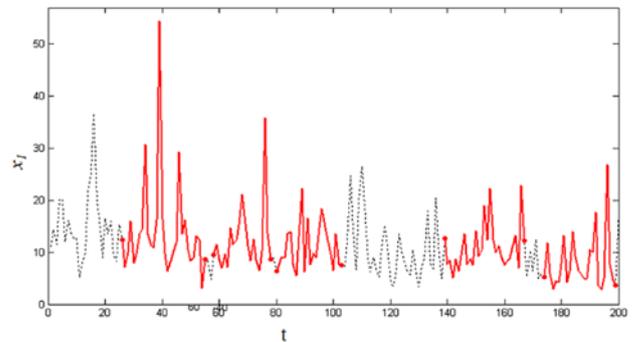

Fig. 4. Almost symmetrical parts for example 2.

The resulting model, allowing symmetry group are of the form:

$$\mathbf{x}(t+1) = A\mathbf{x}(t) + \Psi_o(t),$$
$$y = C\mathbf{x},$$

$$A = \begin{bmatrix} 0.9413 & -0.1805 & 0.1164 & -0.0295 \\ -0.0545 & 0.8226 & 0.1622 & 0.1056 \\ 0.0014 & -0.0105 & -0.4455 & 0.8471 \\ -0.0062 & 0.0341 & -0.8860 & -0.5404 \end{bmatrix},$$

$$\Psi_0 = \begin{bmatrix} 0.0399 \\ 0.0463 \\ -0.4848 \\ -0.1851 \end{bmatrix} \left( \exp(t^{0.0001}) \sin(t^{0.4}) \right),$$

$$C = 10^4 \begin{bmatrix} 2.1037 & -0.0124 & 0.1202 & -0.0302 \end{bmatrix}.$$

## IV. Generation of robust chaos

In the recent literature is widely discussed problem generating the so-called robust chaos [6, 7]. We say a chaotic attractor is robust if, for its parameter values there exists a neighborhood in the parameter space with no periodic attractor and the chaotic attractor is unique in that neighborhood [8].

This refers to such a chaotic dynamics, in which variations in the parameters does not arise "windows periodicity", and the dependence of the largest Lyapunov exponent is a smooth function in a wide range. Emphasizes that it is desirable to have such chaos for applications including communications circuits, a random number generator and coding systems information.

To implement this type offer the chaos system to use, are included in the composition elements, having the characteristics as functions kinked. In practice unattainable ideal break windows and completely eliminate periodicity in this way is problematic.

In systems with uniformly hyperbolic attractors mentioned properties of chaos act as a natural attribute of these systems due to their inherent structural stability [9].

As is known, the robustness of the generated chaotic signal cannot be securely smooth function. However, it can be provided in a piecewise smooth maps [8].

The resulting models can generate robust chaotic signal, based on the use of additional functions in the model:

$$\mathbf{x}(t+1) = A\mathbf{x}(t) + \Psi_o(t)u(t),$$
$$y = C\mathbf{x},$$

where $u(t)$ — piecewise continuous function. For practical examples, we used a linear function of the form $u = pt + q_i$.

Comparison of the generated chaotic signal with the reference near shown in Fig. 5.

## V. Conclusion

The resulting models can simulate the chaos with desired properties. Possible view of the attractor set, view and extent permitted symmetries symmetry violation.

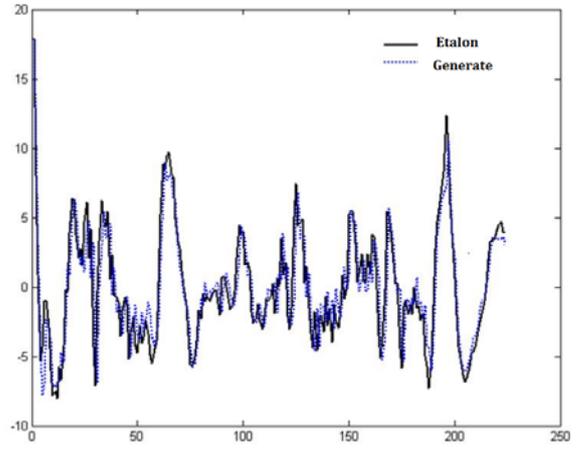

Fig. 5. Chaotic series generated by the method of symmetry violation

In conclusion, we note that on the basis of a pilot study found that the use of these models reduce demand on the sensitivity of the method of least squares for parametric identification matrices.